\RequirePackage{snapshot}
\documentclass[aps,pra,twocolumn,superscriptaddress,nofootinbib,showpacs]{revtex4-1}

\usepackage[english]{babel}
\usepackage{amsmath}
\usepackage{amsfonts}
\usepackage{amssymb}
\usepackage{amsthm}
\usepackage{graphicx}
\usepackage{bm}

\begin{document}

\title{Pulse shaping in the terahertz frequency range for the control of photo-excited carriers in graphene}

\author{Denis Gagnon}
\affiliation{D\'epartement de Chimie, Universit\'e de Montr\'eal, Montr\'eal, Qu\'ebec, Canada}
\author{Joey Dumont}
\affiliation{Universit\'e du Qu\'ebec, INRS--\'Energie, Mat\'eriaux et T\'el\'ecommunications, Varennes, Qu\'ebec, Canada}
\author{Fran\c{c}ois Fillion-Gourdeau}
\affiliation{Universit\'e du Qu\'ebec, INRS--\'Energie, Mat\'eriaux et T\'el\'ecommunications, Varennes, Qu\'ebec, Canada}
\affiliation{Institute for Quantum Computing, University of Waterloo, Waterloo, Ontario, Canada}
\author{Steve MacLean}
\email{steve.maclean@uwaterloo.ca}
\affiliation{Universit\'e du Qu\'ebec, INRS--\'Energie, Mat\'eriaux et T\'el\'ecommunications, Varennes, Qu\'ebec, Canada}
\affiliation{Institute for Quantum Computing, University of Waterloo, Waterloo, Ontario, Canada}

\date{\today}
 
\begin{abstract}
	
	The shape of a few-cycle terahertz (THz) laser pulse can be optimized to provide control over conduction band populations in graphene.
	To demonstrate this control in a theoretical way, a spectral parametrization of the driving pulse using $B$-splines is used in order to obtain experimentally realistic pulses of bandwidth $\sim$ 30 THz.
	Optimization of the spectral shape is performed via differential evolution, using the $B$-splines expansion coefficients as decision variables.
	Numerical results show the possibility of changing the carrier density in graphene by a factor of 4 for a fixed pulse energy.
	In addition, we show that it is possible to selectively suppress or enhance multi-photon absorption features by optimizing over narrow windows in reciprocal space.
	The application of pulse shaping to the control of scattering mechanisms in graphene is also discussed.
	
\end{abstract}

\maketitle

\section{Introduction}
The electronic band structure of graphene, characterized by a linear dispersion relation, makes it an ideal material for the study of the fundamentals of light-matter interactions \cite{Geim2007, CastroNeto2009}.
In fact, because of the small value of its Fermi velocity and the absence of a bandgap in monolayer graphene, multiphoton transitions in the strong-field regime correspond to accessible laser frequencies, i.e. in the visible range or smaller.
As a direct result of these properties, graphene is sensitive to the temporal shape of strong-field, few-cycle pulses \cite{Lefebvre2017}.
This sensitivity has led to many applications of laser-driven graphene such as controlling directional photo-currents \cite{higuchi2017light}, graphene polarizers \cite{bao2011broadband,Farmani2018b} and giant lateral shifts \cite{Farmani2018a,Farmani2017}.

The research field of ``high-field physics'' in graphene has gained momentum with the advent of intense terahertz (THz) radiation sources \cite{Razavipour2015, McGouran2016}: short pulses with peak electric fields in the kV/cm range are now commonplace \cite{Ropagnol2013, Hafez2016}, with some leading-edge sources even reaching the MV/cm threshold \cite{Ishikawa2013}.
As an illustration, THz radiation can lead to high-harmonic radiation when used to pump a graphene mono-layer \cite{Al-Naib2014}.
In addition, the ultra-fast carrier dynamics in graphene induced by short THz pulses can be probed using techniques such as angle-resolved photoemission spectroscopy (ARPES) \cite{Ishikawa2013, Sentef2015, Kelardeh2016, Aeschlimann2017}.

The result of momentum-resolved experiments with graphene has been the subject of several articles in recent years, most of which consider the effect of a short optical or THz pulse on the conduction band population in reciprocal space \cite{Ishikawa2013, Kelardeh2015, Kelardeh2016, Fillion-Gourdeau2015, Fillion-Gourdeau2016, Lefebvre2017}.
Theoretical investigations have revealed that the details of the temporal pulse shape, for instance the carrier-envelope phase, can have a manifest impact on the momentum space patterns \cite{Ishikawa2013, Lefebvre2017}.
In this article, we consider the \emph{inverse} problem, namely finding the temporal pulse shape which minimizes the photo-induced carrier density in graphene over a pre-defined momentum window.
The numerical solution of this problem is performed using differential evolution (DE), a high-level optimization algorithm.
Pulse shaping problems are also encountered in several other areas of physics, for instance atomic and molecular physics, quantum electrodynamics (QED) and quantum information, and are tackled using various high-level techniques \cite{Chu2001, Christov2001, Balogh2014, Hebenstreit2016, Fillion-Gourdeau2017}.
The results of this article may thus find application in other subfields of physics besides laser-irradiated graphene and related Dirac materials.
Control over THz pulse shapes through spectral amplitudes or phases has been performed recently using several techniques such as photoexcited semiconductors \cite{doi:10.1063/1.4907635}, dynamic waveguides \cite{Gingras:18} and plasmonic metasurfaces \cite{Rahimi:16}. 

In this work, optimization constraints on the THz spectrum are chosen to obtain short pulses that can be generated experimentally.
To achieve this theoretically, the pulse spectrum is parametrized using a $B$-spline polynomial basis, and the corresponding expansion coefficients are used as inputs of the DE solver following a procedure described in Ref. \cite{Fillion-Gourdeau2017} for the optimization of Schwinger's mechanism in QED.
This parametrization ensures a certain level of smoothness in the spectrum of the resulting THz pulses, and results in a low-dimensional search space, thus speeding up optimization runs.
Other than this choice of a $B$-spline basis, few assumptions are made as to the details of the final pulse. 
The main finding of the article is that the photo-induced carrier density in graphene can be varied by a factor of 4 for a fixed THz fluence, leading to the possibility of using spectral optimization as a control knob over scattering mechanisms in Dirac materials.
In addition, we are able to selectively suppress or enhance multiphoton absorption features by using the pulse fluence as an additional control parameter in optimization calculations.

This article is organized as follows.
The pulse shaping problem central to this work is presented in section \ref{sec:problem}.
Specifically, we discuss the $B$-spline parametrization used for optimization calculations (Sec. \ref{sec:param}), as well as the optimization solver  itself (Sec. \ref{sec:de}).
Two different problems are then discussed (Sec. \ref{sec:results}) and a summary is given (Sec. \ref{sec:summary}).

\section{Problem definition}\label{sec:problem}
Consider Dirac fermions in a graphene mono-layer in the presence of a linearly polarized electric field, which is uniform in space and parallel to the graphene plane.
Assuming that electron-electron interactions and carrier relaxation can be neglected (the validity of these assumptions is discussed in Sec.  \ref{sec:cfluence}), the fermion dynamics are governed by the time-dependent Dirac equation (TDDE) in reciprocal space, or $\mathbf{p}$-space (we use units such that $\hbar = 1$):
\begin{equation}\label{eq:Dirac}
i\partial_t \psi_{s,\textbf{K}_{\pm}}(t,\textbf{p}) = H_{\textbf{K}_{\pm}} \psi_{s,\textbf{K}_{\pm}}(t,\textbf{p}),
\end{equation}
where  $\psi_{s,\textbf{K}_{\pm}}(t_f,\textbf{p})$ is the wavefunction, $s$ is the physical spin of the electron and $\textbf{K}_{\pm}$ are non-equivalent Dirac points, corresponding to the valley pseudospin.
The graphene Hamiltonian is characterized by the usual linear dispersion relation and the absence of a mass term:
\begin{equation}\label{eq:Hamiltonian}
H_{\textbf{K}_{\pm}} (t,\textbf{p}) = \pm v_{F} \bm{\alpha} \cdot \big(\textbf{p}+e\textbf{A}(t) \big)
\end{equation}
where $\textbf{A}(t)$ is the time-dependent vector potential, $v_F$ is the graphene Fermi velocity and $\bm{\alpha}$ is the dyad of Pauli matrices in the space of the two sublattices of graphene \cite{Rodionov2016}.
In this work, we consider the problem wherein a short THz pulse described by the vector potential $\textbf{A}(t)$ drives non-adiabatic transitions from the valence to the conduction band of graphene.
A non-adiabatic transition from the valence to the conduction band corresponds to a ``flip'' of the sublattice pseudospin \cite{Song2015, Aeschlimann2017}.
This dynamical interband process may result in a finite conduction band population after the passage of the pulse \cite{Kelardeh2015, Lefebvre2017}, a population which can then be probed using time-resolved ARPES \cite{Ishikawa2013, Sentef2015, Kelardeh2016, Aeschlimann2017}.
An alternative way of describing this physical process is that, for a given quasiparticle momentum $\textbf{p}$, the valence and conduction band of graphene behave like a driven two-level atom.
When the driving amplitude is large, as is the case for a strong THz pulse, multiphoton processes become relevant \cite{Shevchenko20101}.

In this article, the physical observable used in optimization calculations is the electron momentum density (EMD), similar to other theoretical studies of laser-irradiated graphene \cite{Ishikawa2013, Kelardeh2015, Fillion-Gourdeau2016, Lefebvre2017}.
This observable is calculated as follows: for a given value of $\textbf{p}$, a free negative energy state is ``prepared'' and propagated numerically up to the final time $t_f$ (i.e. when the vector potential does not vary anymore) using a split-operator decomposition described in Refs. \cite{Fillion-Gourdeau2016, Fillion-Gourdeau2017a}.
The EMD is then obtained by projecting the numerically computed wavefunction onto a free positive energy state of the TDDE  $u_{s,\textbf{K}_{\pm}}^{\mathrm{out}\dagger}$ :
\begin{equation}\label{eq:pairDensity}
f(t_f, \mathbf{p}) = \frac{1}{2 \epsilon_{\textbf{p}}^{\mathrm{out}}2 \epsilon_{\textbf{p}}^{\mathrm{in}}}\arrowvert u_{s,\textbf{K}_{\pm}}^{\mathrm{out}\dagger}(\textbf{p}) \psi_{s,\textbf{K}_{\pm}}^{\mathrm{out}}(t_f,\textbf{p})\arrowvert ^2,
\end{equation}
where $\epsilon_{\textbf{p}}^{\mathrm{in,out}}$ are the asymptotic eigen-energies.
This observable is equal to the photo-induced pseudospin flip probability, in other words the induced carrier density \cite{Fillion-Gourdeau2016}.
To obtain the total carrier density, this number may be multiplied by 4 to take into account the physical spin and valley pseudospin degeneracies.
We assume undoped graphene in this article in other words a Fermi energy equal to zero.
This implies that all transitions between the hole-like states with momentum $-\textbf{p}$ and electron like-states with momentum $\textbf{p}$ are allowed.
Control of the Fermi energy in graphene at THz frequencies can be achieved through gating of the sample \cite{Razavipour2015}.
We also assume that the sample is at absolute zero during the interaction time with the pulse. 
Thus, our calculations do not include the effect of electron-electron interactions \cite{DasSarma2011} as well as thermal intraband transitions that can occur at non-zero temperatures \cite{Farmani2018b}.
The validity of these assumptions is discussed in section \ref{sec:cfluence}.

The basic procedure used in this work is as follows: an objective function which depends on the physical observable $f(t_f, \mathbf{p}) $ -- \eqref{eq:pairDensity} -- is defined.
Then, decision variables related to an appropriate parametrization of the vector potential $\textbf{A}(t)$ entering in \eqref{eq:Hamiltonian} are chosen.
These decision variables are used as inputs of the DE solver, and the vector potential $\textbf{A}(t)$ which minimizes the objective function is found numerically.
The parametrization of the vector potential is detailed in subsection \ref{sec:param}, while the specifics of the DE solver are presented in subsection \ref{sec:de}.

\subsection{Terahertz field parametrization}\label{sec:param}
\begin{figure}
	\centering
	\includegraphics[width=0.4\textwidth]{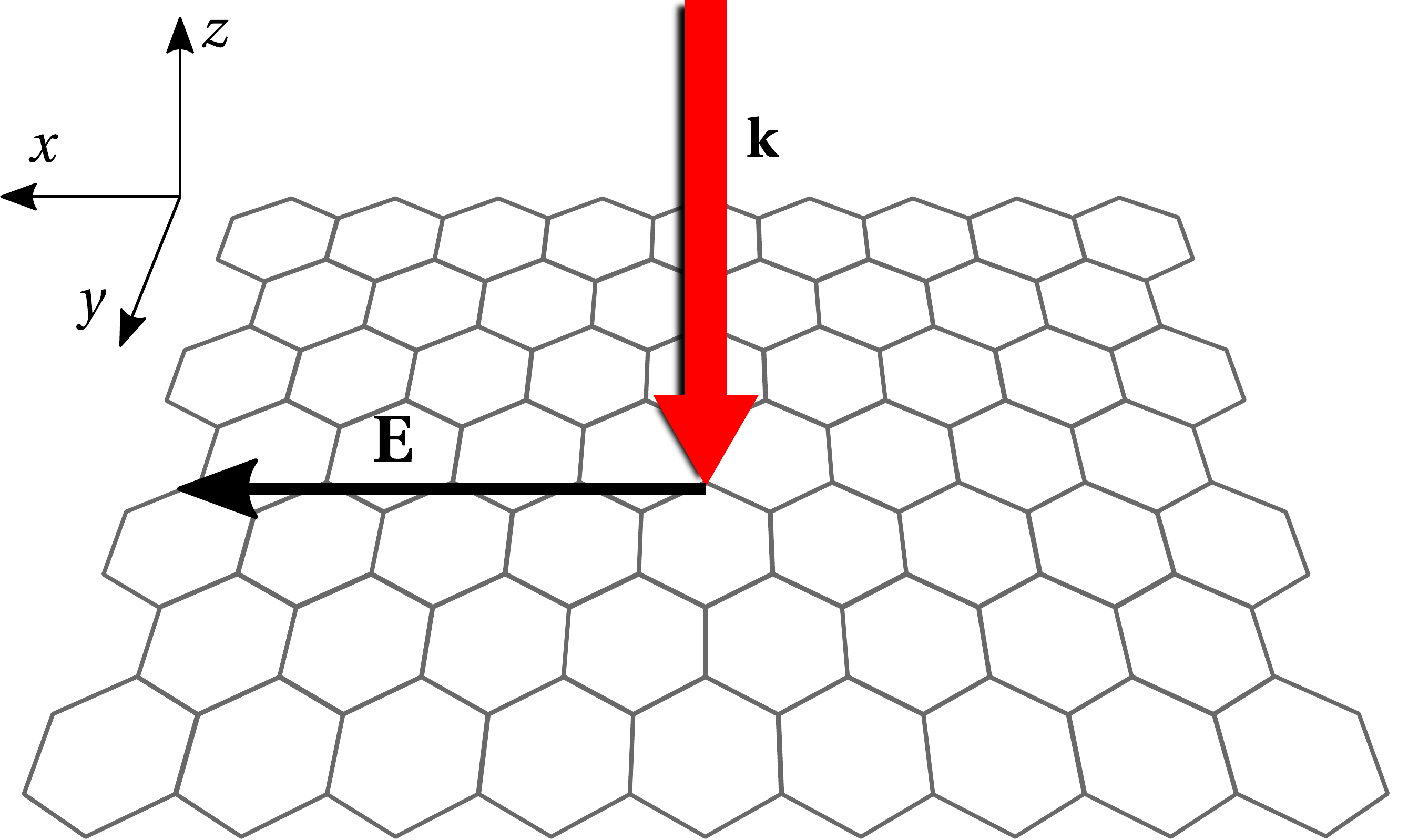}
	\caption{Graphene sheet irradiated by an linearly polarized excitation.
		The driving field $E(t)$ is polarized in the $x$-direction, corresponding to normal incidence.}
	\label{fig:lightmatter}
\end{figure}

In this article, linearly polarized oscillating fields in the THz frequency range are considered [see Fig. \ref{fig:lightmatter} for a schematic].
The fields vary only in the time-domain, and are assumed homogeneous in space.
Starting from the definition of the velocity gauge, one can write
\begin{equation}\label{eq:gauge}
\mathbf{E}(t) = - \dfrac{d \mathbf{A}(t)}{dt} = E(t) \hat{\mathbf{e}}_x.
\end{equation}
The time-dependent field is fully characterized by its spectral density $\tilde{E}(\omega)$.
Following the approach described in Ref. \cite{Fillion-Gourdeau2017}, the density is expanded over a polynomial basis as
\begin{equation}\label{eq:density}
\tilde{E}(\omega) = \sum_{i=1}^{N_{s}} a_i B_{i}(\omega)
\end{equation}
where $N_s$ is the number of basis elements, $a_i$ are expansion coefficients and $B_i$ are the basis elements.
As will be described in subsection \ref{sec:de}, the expansion coefficients $a_i$ are chosen as inputs of the optimization solver.
This choice is motivated by the fact that it allows one to search a low-dimensional parameter space, and then ``oversample'' the optimized spectral density described by \eqref{eq:density}, thus yielding a smooth, physically realistic THz pulse.
The choice of basis elements is not unique.
In this work, $B$-spline polynomials of order $k$ are used by design:
\begin{equation}
B_i(\omega) = b_i^{(k)}(\omega).
\end{equation}
A detailed description of $B$-splines can be found in Ref. \cite{Bachau2001}, while their application to pulse shaping problems is described in Ref. \cite{Fillion-Gourdeau2017}.
In a nutshell, they are favored over other orthogonal polynomials because of their compact support, positive definiteness, the fact that they are easy to generate numerically and the ease of managing boundary conditions with $B$-splines.

$B$-splines are fully determined by their polynomial order, $k$, and a knot vector $(\omega_{i})_{i=1,\ldots,N_{s}+k}$ according to the recurrence relation \cite{Bachau2001, DeBoor1978}
\begin{align}
b_{i}^{(k)}(\omega)&= \frac{\omega-\omega_{i}}{\omega_{i+k-1} - \omega_{i}}b^{(k-1)}_{i}(\omega) + \frac{\omega_{i+k} -\omega}{\omega_{i+k}-\omega_{i+1}} b^{(k-1)}_{i+1}(\omega).
\end{align}
The following initial condition is used to generate $B$-spline coefficients
\begin{equation}
b_{i}^{(1)}(\omega)=\begin{cases} 1 & \mbox{for} \quad \omega_{i} \leq \omega < \omega_{i+1}\\ 0 & \mbox{otherwise}\end{cases}.
\end{equation}
The number of knots at a given frequency determines the continuity condition at that point.
In this article, we use the standard choice with knots of multiplicity $k$ at the endpoints $\omega_{\rm{min}}$ and $\omega_{\rm{max}}$, and knots of multiplicity 1 (simple knots) at the interior points \cite{Bachau2001}
\begin{align}\label{eq:knot}
\omega_{\rm{min}}  = \omega_{1}  = \ldots & = \omega_{k} < \cdots  \nonumber \\ 
< \omega_{k+n-1} & = \ldots =\omega_{2k+n-2} = \omega_{\rm{max}} \, ,
\end{align}
where $n$ is the number of breakpoints and $2k+n-2$ is the number of knot points.
These two quantities are related to the total number of $B$-splines as $N_{s} = n+k-2$. 
The bandwidth of the parametrized spectrum is fixed by the endpoints $\omega_{\mathrm{min}}$ and $\omega_{\mathrm{max}}$.
Outside of this interval, the spectral density is zero by definition.

Boundary conditions can be enforced at the edges of the spectrum by simply removing functions from the basis set, \eqref{eq:density}, although this is not necessary in principle \cite{Fillion-Gourdeau2017}.
This approach is equivalent to changing the multiciplicity of knots at the endpoints of the knot vector, \eqref{eq:knot}.
At $\omega = \omega_{\mathrm{min}}$, we impose that the spectral density is zero, but do not enforce continuity.
The corresponding condition is
\begin{equation}\label{eq:smooth0}
a_1 = 0.
\end{equation}
since $B_1$ it is the only non-zero spline at $\omega = \omega_{\mathrm{min}}$ \cite{Bachau2001}.
At $\omega = \omega_{\mathrm{max}}$, we require that the spectrum is zero, but that it decreases to zero in a smooth manner.
Since $B_{N_s - 2}$ is the last spline in the basis set which tends to zero smoothly, the condition used to enforce this behavior is 
\begin{equation}\label{eq:smooth1}
a_{N_s} = a_{N_s - 1} = 0.
\end{equation}
We have found this choice to yield the most physically realistic THz pulse spectrum possible, that is a spectral density that is continuous everywhere except at $\omega = \omega_{\mathrm{min}}$ where it may reach zero in a nearly discontinuous way \cite{Lee2002, Hafez2016}.

Given a $B$-spline parametrization, one can determine the temporal field shape by resampling the spectrum at equidistant angular frequencies $\Omega_j = (N_\mathrm{min} + j) \Delta \omega$ where $j \in [0,j_\mathrm{max}] \subset \mathbb{N}$.
A given spectral resolution $\Delta \omega$ results in a periodicity $T = 2 \pi / \Delta \omega$ of the temporal field profile.
The bounds on the sampled spectrum are thus $\Omega_j \in [\omega_\mathrm{min}, \omega_\mathrm{max}]$ where $\omega_\mathrm{min} = N_\mathrm{min} \Delta \omega$ and $\omega_\mathrm{max} = (N_\mathrm{min} + j_\mathrm{max})\Delta \omega$.
Using this sampling, the temporal shape of the applied electric field is given by a superposition of $N+1$ harmonic modes:
\begin{equation}\label{eq:temporal}
E(t) = \sum_{j=0}^N \tilde{E}(\Omega_j) \cos (\Omega_j t).
\end{equation}
Since the spectral phase is set to zero in this work, \eqref{eq:temporal} describes a field which reaches zero at the endpoints of the temporal interval, i.e at $t=\pm T/2$.
The corresponding vector potential can be calculated directly by combining \eqref{eq:gauge} and \eqref{eq:temporal}, yielding
\begin{equation}
A(t) = - \sum_{j=0}^N  \tilde{E}(\Omega_j) \frac{\sin (\Omega_j t)}{\Omega_j},
\end{equation}
with $\textbf{A}(t) = A(t) \hat{\mathbf{e}}_x$.
In actuality, before computing the vector potential $A(t)$ from the value of the applied electric field $E(t)$, the field expression \eqref{eq:temporal} is renormalized to obtain a given value of the pulse fluence, or energy density:
\begin{equation}
U=\frac{c \epsilon_0}{2} \int_{-\infty}^{\infty}{\mkern-10mu dt\, E^2(t)}.
\end{equation}
As described further in Sec. \ref{sec:results}, the fluence of the pulse $U$ may remain fixed in calculations or be used as a decision variable.
Fluences around $10^{-4}$ J/cm$^2$ are used in this article.

The $B$-spline parametrization presented in this work, \eqref{eq:density}, can be extended to treat circular polarizations \cite{Fillion-Gourdeau2017} or other simple electric field configurations.
However, we assume an homogeneous time-dependent electric field in the form of \eqref{eq:temporal} to solve the optimization problem in a reasonable time, and because this simple model captures the main physical processes of laser-graphene interactions in the case of normal incidence [see Fig. 1].

\subsection{Optimization solver}\label{sec:de}
The problem considered in this article consists in finding a THz pulse shape that results in a given carrier density over a pre-defined quasiparticle momentum range.
As stated earlier, the decision variables are related to the pulse fluence as well as the spectral parametrization via $B$-splines, \eqref{eq:density}.
In mathematical terms, this can be written as a minimization problem
\begin{equation}\label{eq:problem}
\tilde{J} = \min_{\vec{X} \in \mathbb{R}^N} J(\vec{X})
\end{equation}
where $\tilde{J}$ is a minimum of the objective function $J$ in parameter space, $\vec{X} = \big\lbrace X_1, X_2 , \cdots , X_N \big\rbrace$ is a decision vector composed of $N$ decision variables (the choice of these variables will be detailed in section \ref{sec:results}).
The objective function $J(\vec{X})$ is defined in terms of the integrated carrier density in graphene:
\begin{equation}
\label{eq:opt_prob}
J(\vec{X}) =\pm \int_{\mathcal{D}_{\mathbf{p}}}{\mkern-10mu d^2 p\,f(t_f,\mathbf{p})}
\end{equation} 
where $\mathcal{D}_{\mathbf{p}}$ is a pre-defined integration range in momentum space and $f$ is defined in \eqref{eq:pairDensity}.
The sign in front of the integral in \eqref{eq:opt_prob} can be chosen positive if one wishes to minimize the carrier density, and negative if one wishes to maximize it.

\begin{figure}
	\centering
	\includegraphics[width=\columnwidth]{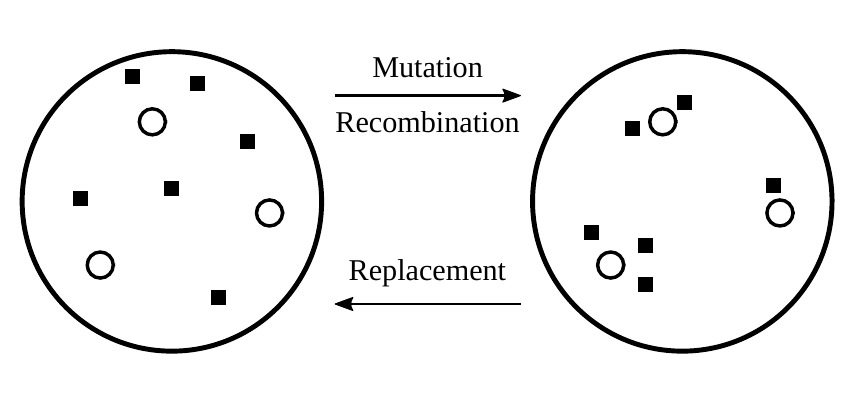}
	\caption{High-level template of DE. Different initial conditions or ``individuals'' (black squares) are generated randomly in the parameter space (enclosing circle). This population of individuals is iteratively improved using successive applications of mutation, crossover and selection operators. Every generation features individuals successively closer to local minima (small circles) of the problem. Inspired by \cite{Talbi2009}.}
	\label{fig:de}
\end{figure}

The search space of the optimization problem \eqref{eq:problem} is a $N$-dimensional hypercube, where $N$ is the number of decision variables.
We start from the hypothesis that the effect of individual decision variables on \eqref{eq:opt_prob} can not be readily isolated, in other words that the search landscape is non-linear.
Accordingly, we use DE, an algorithm that has been successfully applied to the suppression of multiphoton resonances in driven two-level systems \cite{Gagnon2017b} and to the maximization of pair production in QED \cite{Fillion-Gourdeau2017}. 
The algorithm begins with a randomly chosen set of initial guesses called ``individuals', then mutation and recombination operators direct the search towards good solutions using vector differences \cite{Talbi2009}.
It is similar to other population-based algorithms that are often used to tackle pulse shaping problems, for example the genetic algorithm (GA) \cite{Chu2001, Christov2001, Balogh2014, Hebenstreit2016}.
To be concise, DE is a population based, evolutionary optimization algorithm targeted at continuous parameter spaces [see Fig. \ref{fig:de} for visualization].

Several variants of DE exist, most of which differ only in the specifics of how individuals are updated (mutations and recombinations).
The variant used in this article (DE/rand/1/bin in DE notation \cite{Talbi2009}) is the same as in Ref. \cite{Fillion-Gourdeau2017}, thus we refer the interested reader to this article for details.
In this work we make use of the \textsc{Pagmo} optimization library \cite{izzo2012pygmo} which contains the aforementioned variant of the DE solver \cite{Storn1997,das2011differential}.

To conclude this description of pulse shaping calculations, let us summarize the basic procedure used to generate numerical results in this article.
\begin{enumerate}
	\item A momentum range $\mathcal{D}_{\mathbf{p}}$ and the objective function $J(\vec{X})$, \eqref{eq:opt_prob}, are defined, either with respect to maximization or minimization of the induced carrier density.
	\item Optimization variables are chosen. In this work we use the coefficients of the $B$-spline parametrization of the spectrum, \eqref{eq:density}, and/or the pulse fluence.
	\item The chosen decision variables, which are stored in a vector $\vec{X}$ are used as inputs of the DE solver: 
	an initial population of solutions is chosen randomly, and is evolved via DE for a fixed number of iterations (often called ``generations'').
	For every individual and generation, the objective function $J(\vec{X})$ is evaluated by evolving the TDDE and computing the subsequent density, \eqref{eq:pairDensity}.
	This step returns a possible ``optimum'' of the problem, that is the value of the decision vector $\vec{X}$ which best optimizes the objective function.
\end{enumerate}
Step 3 may be repeated using different random initial populations until a satisfactory solution is found.

\section{Results and discussion}
\label{sec:results}

As stated earlier, the goal of the calculations presented in this work is to optimize the spectral content of a THz pulse for the suppression or enhancement of the induced carrier density over a pre-defined momentum range.
For comparison purposes, the THz bandwidth and the $B$-splines parametrization of the spectrum [see \eqref{eq:density}] are the same throughout the article [see Table \ref{tab:pulse} for details].
The bandwidth of the pulse is set to 30 THz, and the number of $B$-splines is fixed to 10.
Keeping in mind that some $B$-spline coefficients are fixed to zero [see \eqref{eq:smooth0} and \eqref{eq:smooth1}], the number of decision variables accessible to the DE solver is 7 or 8, depending on if the pulse fluence is kept constant or is varied.
Two different situations are considered in this article: the case of fixed pulse fluence (sec. \ref{sec:cfluence}) and the case of a variable pulse fluence (sec. \ref{sec:fluence}).
We detail the choice and bounds of decision variables in each of these sections, then show optimized pulse shapes and the resulting carrier density in reciprocal space.

\begin{table}[t!]
	\begin{tabular}{lc}
		\hline \hline
		Simulation parameter & Value \\
		\hline
		Minimum frequency ($\nu_{\mathrm{min}} = \omega_{\mathrm{min}}/2\pi$)& 1 THz \\
		Maximum frequency ($\nu_{\mathrm{max}} = \omega_{\mathrm{max}}/2\pi$)& 30 THz \\
		Total length ($T=2 \pi/\Delta\omega$) & 2 ps \\
		Number of $B$-splines ($N_s$) & 10 \\
		$B$-spline order ($k$) & 3 \\
		\hline \hline
	\end{tabular}
	\caption{Simulation parameters used in pulse shaping calculations throughout this work. The pulse fluence is not fixed in advance as it can be used as a decision variable.}
	\label{tab:pulse}
\end{table}

\subsection{Carrier density optimization}\label{sec:cfluence}
In this section, we consider minimization and maximization of the carrier density over a wide momentum range for a fixed pulse fluence,
The variables composing the decision vector $\vec{X}$ are selected as the non-zero coefficients in \eqref{eq:density}, that is
\begin{equation}
X_n = a_{n+1},
\end{equation}
with the number of decision variables set to $N = N_s -3$, i.e. $N=7$. 
Bounds on $\vec{X}$ are fixed as
\begin{equation}
\begin{aligned}
\min(X_n) & = 0, \\
\max(X_n) & = 1,
\end{aligned}
\end{equation}
and the harmonic superposition, \eqref{eq:temporal} is re-normalized at every objective function evaluation such that the applied THz field has a constant fluence from iteration to iteration.

\begin{figure}
	\includegraphics[width=0.5\textwidth]{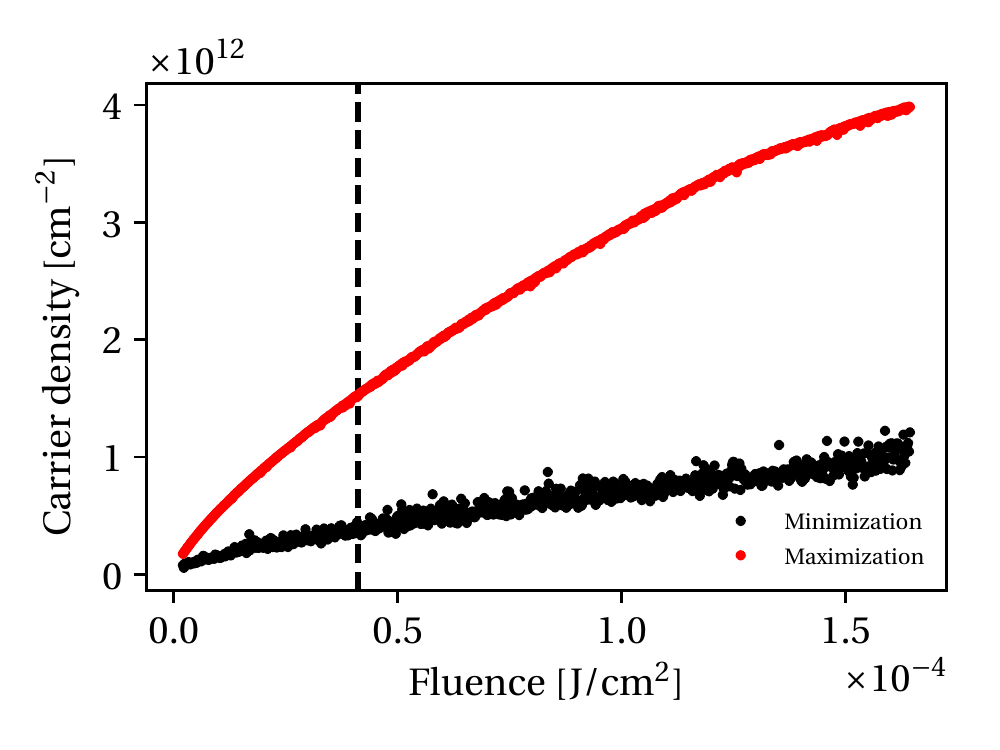}
	\caption{Values of the optimized carrier density for a range of THz pulse fluences. Every point on the plot corresponds to an optimized configuration found via DE, with 300 generations and a population size of 8.
		The problem parameters are specified in Table \ref{tab:pulse} and the optimization window is $\mathcal{D}_{\mathbf{p}}$ = $\lbrace 0 \leq k_x \leq 2.6 \times 10^{-1} \mbox{ \AA}^{-1},  0 \leq k_y \leq 7.2 \times 10^{-2} \mbox{ \AA}^{-1} \rbrace$. The valley and physical spin degeneracies are taken into account in the carrier density values. 
		The dashed line indicates the fluence value for the results shown in Figs. \ref{fig:maximization}--\ref{fig:minimization}.}
	\label{fig:fluence_plot}
\end{figure}

The reciprocal space window over which optimization takes place is chosen as $\mathcal{D}_{\mathbf{p}}$ = $\lbrace 0 \leq k_x \leq 2.6 \times 10^{-1} \mbox{ \AA}^{-1},  0 \leq k_y \leq 7.2 \times 10^{-2} \mbox{ \AA}^{-1} \rbrace$.
This 2D window covers all the quasi-particle states that can become excited via a THz pulse for the parameters found in Table \ref{tab:pulse} and a fluence below $2\times 10^{-4}$ J/cm$^2$.
For a given fluence value, minimization via DE is carried out for 300 generations using a randomly initialized population of 8 individuals. 
This calculation is repeated for maximization using the same optimization parameters.
Minimization and maximization calculations are performed for a range of fluence values ranging from 0.02 to $1.64 \times 10^{-4}$ J/cm$^2$.
The results of this parameter sweep are shown in Fig. \ref{fig:fluence_plot}.

The spectral content of the optimized pulse can be used to control the photo-induced carrier density, as shown by the optimization results in Fig. \ref{fig:fluence_plot}.
As a matter of fact, for pulse fluences around $1.5 \times 10^{-4}$ J/cm$^2$, the maximum achievable carrier density can be as high as 4 times the minimum achievable carrier density.
We stress the fact that only the spectral content varies for pulses of equal fluence: the total energy deposited in the graphene sample remains the same.

\begin{figure}
	\includegraphics[width=0.5\textwidth]{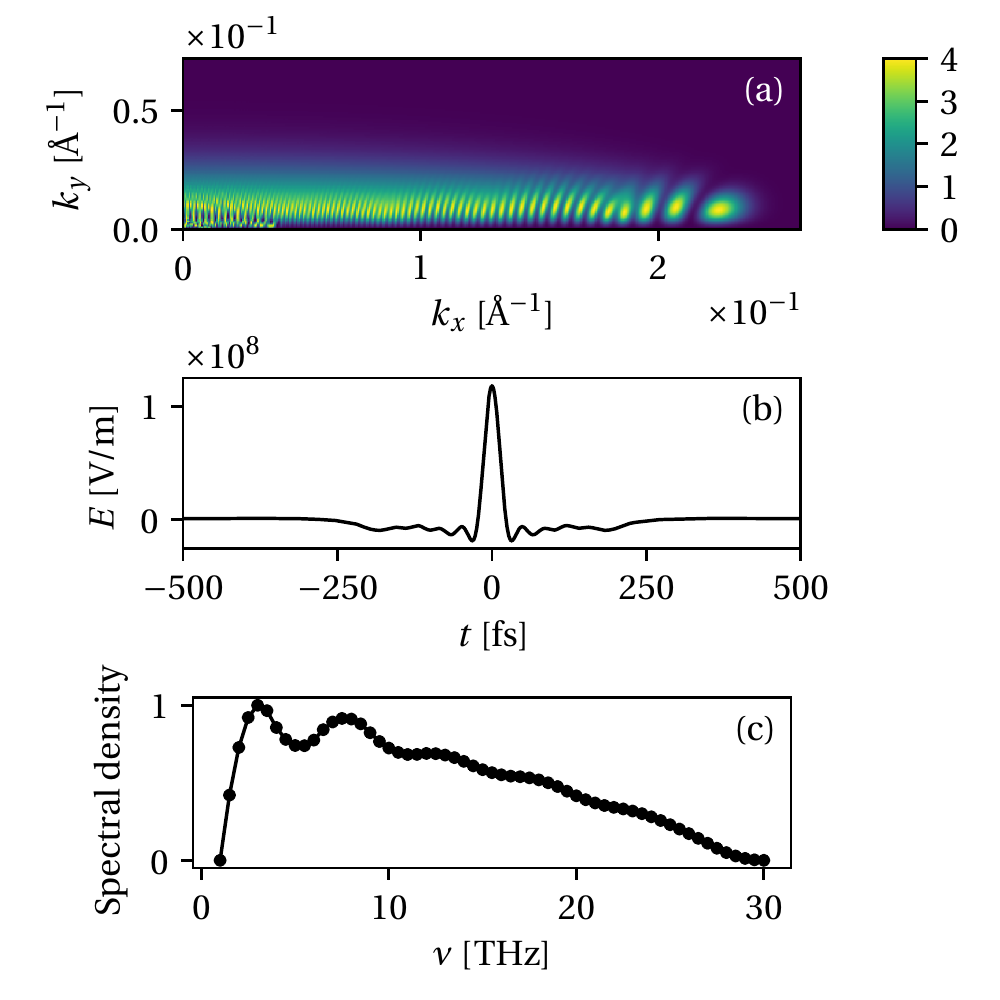}
	\caption{Best found solution for the maximization of the induced carrier density at a THz fluence of $U=4.11 \times 10^{-5}$ J/cm$^2$. 
		(a) Optimized density in momentum space. (b) Associated temporal pulse shape. The peak field is determined by the pulse fluence. (c) Optimized spectral density from which the temporal shape is obtained (arbitrary units). The markers in (c) are re-sampling points used to calculate the temporal pulse shape, different from the number of $B$-splines which is $N_s = 10$.}\label{fig:maximization}
\end{figure}

\begin{figure}
	\includegraphics[width=0.5\textwidth]{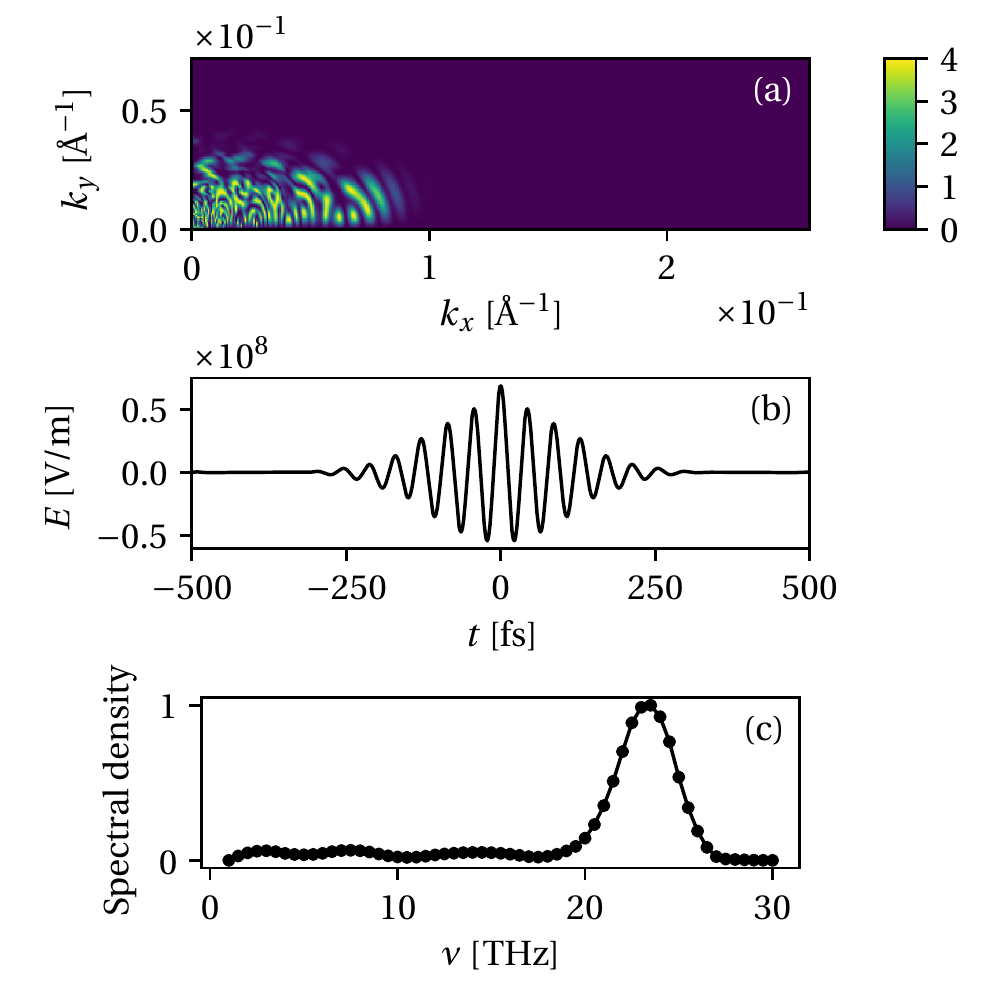}
	\caption{Best found solution for the minimization of the induced carrier density at a THz fluence of $U=4.11 \times 10^{-5}$ J/cm$^2$. 
		(a) Optimized density in momentum space. (b) Associated temporal pulse shape. The peak field is determined by the pulse fluence. (c) Optimized spectral density from which the temporal shape is obtained (arbitrary units). The markers in (c) are the sampling points used to calculate the temporal pulse shape, different from the number of $B$-splines which is $N_s = 10$.}
	\label{fig:minimization}
\end{figure}

Let us examine in more detail the pulse shapes resulting in maximization/minimization of the carrier density for a fluence of $4.11 \times 10^{-5}$ J/cm$^2$, indicated by a dashed line in Fig. \ref{fig:fluence_plot}.
The optimized spectral pulse shapes, as well as the associated temporal shapes and momentum-resolved carrier density, are shown in Fig. \ref{fig:maximization} (maximization) and in Fig. \ref{fig:minimization} (minimization).
For this fluence value, carriers are created only within the optimization window $\mathcal{D}_{\mathbf{p}}$, with a zero probability of photo-excitation outside of the window (see Figs. \ref{fig:maximization}a and \ref{fig:minimization}a).
As can be seen by comparing Fig. \ref{fig:maximization}b and Fig. \ref{fig:minimization}b, the DE solver tends to converge towards few-cycle THz pulses when asked to maximize the carrier density, and to multi-cycle pulses when asked to minimize.
Further information can be obtained by comparing the optimized spectra in Fig. \ref{fig:maximization}c and Fig. \ref{fig:minimization}c: a broadband spectrum results in more momentum states being excited by the pulse, whereas a narrower spectrum results in less states being excited.
Interestingly, although it results in a complicated momentum-space pattern (Fig. \ref{fig:minimization}a), the spectrum optimized for suppression exhibits a simple shaped peak around 22 THz, which corresponds to the carrier frequency of the temporal signal.
At this point of the discussion, we recall that the way the optimized spectra tend to zero at 1 THz and 30 THz is partly constrained by the $B$-spline parametrization detailed in Sec. \ref{sec:param}.
We also recall that the calculated values of the EMD fall between 0 and 4, because they account for the two-fold physical spin and two-fold valley pseudospin degeneracies in graphene.

The appearance of a large number of oscillations in the momentum-space pattern for a narrow spectrum [see an example in Fig. \ref{fig:minimization}a] can be explained by considering periodic driving of a quantum system through an avoided energy crossing.
This effect, also called Landau-Zener-St\"uckelberg (LZS) interferometry, manifests itself in the form of interference fringes in reciprocal space.
This time-domain quantum interference was detailed in previous publications, both in the generic \cite{Shevchenko20101} and graphene-specific case \cite{Fillion-Gourdeau2016}.
In short, extremal values of the transition probability from the valence to the conduction band correspond to constructive or destructive interference between different quantum pathways.
If the number of optical cycles in the driving pulse increases, so does the number of possible pathways leading to constructive/destructive interference: thus, more oscillations in momentum space are observed \cite{Fillion-Gourdeau2016}. 


The fact that pulse shaping can be used to change the photo-induced carrier density for fluences around $1.5 \times 10^{-4}$ J/cm$^2$ [see Fig. \ref{fig:fluence_plot}] provides a control knob over scattering channels in graphene, in addition to gating.
As described in Ref. \cite{DasSarma2011}, different scattering processes have different functional dependences on the excess carrier density $\bar{n}$ in Dirac materials such as graphene \cite{DasSarma2011}.
At low temperatures, the scattering rate from short-range disorder in graphene is proportional to $\sqrt{\bar{n}}$  .
In contrast, as the carrier density $\bar{n}$ increases, Coulomb interactions become screened, leading to a $1/\sqrt{\bar{n}}$ proportionality for this scattering channel.
The effect of both mechanisms has been experimentally quantified for chemical-vapor-deposited (CVD) graphene in Ref. \cite{Yu2016}.
In short, the results of this section show the potential of THz pulse shaping for increasing or decreasing scattering times in graphene by a factor of $\sim 2$.

Let us conclude this section by a discussion of the validity of the assumptions made in this article from the point of view of scattering channels.
Theoretical studies have shown that carrier relaxation in graphene usually takes place in two steps.
The first step is thermalization due to carrier-carrier and carrier-phonon scattering.
Thermalization takes place on a time-scale of the order of $~50$ fs \cite{Malic2017}.
The second step is carrier cooling, wherein excited carriers transfer their excess energy to the lattice on a time-scale of $\sim$ 1 ps \cite{Malic2017}.
Therefore, to be experimentally useful, the duration of optimized pulses should be sub-picosecond, and care should be taken to subject the graphene layer to environmental conditions such that the thermalization time is increased.
This may include, for instance, performing experiments below 10 K temperature \cite{Al-Naib2014} and embedding the graphene mono-layer in a medium with a sufficiently high dielectric constant \cite{Basov2014}.
The latter precaution aims to ensure that electron-electron interactions are suppressed \cite{Couto2011}.
This suppression could in principle increase the thermalization time of the dynamical system to values over 100 fs, thus enabling the measurement of an anisotropic momentum space distribution with a THz pump beam.
Another possibility for increasing the carrier lifetime beyond 100 fs would be to use $n$-doped graphene \cite{Aeschlimann2017}, although in this case some transitions close to the Dirac point will be forbidden.

\subsection{Control of multiphoton absorption}\label{sec:fluence}

In the previous section, it was shown that the total photo-induced carrier density in graphene could be controlled by the shape of a few-cycle THz pulse.
We now turn our attention to the suppression of specific multiphoton absorption features in the momentum-integrated spectrum of irradiated graphene, i.e. optimization over a narrow momentum window.
As described in Ref. \cite{Gagnon2017b}, selective suppression of multiphoton features is useful not only for the control of scattering mechanisms in Dirac materials, but also from a quantum computing perspective.
In the given reference, the suppression of multiphoton peaks was studied from the point of view of Floquet theory, which applies \emph{stricto sensu} for periodic excitations that exist for all time.
Since the present article is concerned with calculations using the TDDE and pulses of finite duration, it can be viewed as a follow-up on Ref. \cite{Gagnon2017b}. 

As explained in the previous subsection, interference fringes in reciprocal space [see an example in Fig. \ref{fig:maximization}a] can be interpreted in terms of LZS interferometry \cite{Fillion-Gourdeau2016}.
By approaching the LZS problem from the point of view of Floquet theory, it can be shown that the peak amplitude of the applied field has an influence on the suppression of resonances via the phenomenon known as coherent destruction of tunnelling (CDT) \cite{Grifoni1998, Shevchenko20101}.
This can be explained by the fact that a linearly polarized excitation opens a dynamical gap between the valence and conduction band, with the width of the gap a function of the spectral content of the periodic driving pulse.
In optimization calculations presented in this section, we allow the fluence to vary in addition to the spectral content, since this provides an additional control knob for suppressing or enhancing individual peaks in the absorption spectra.

For the remainder of this section, the first $N-1$ variables composing the decision vector $\vec{X}$ are selected as the non-zero coefficients in \eqref{eq:density}, that is $X_n = a_{n+1}$,
with the number of decision variables set to $N - 1 = N_s -3$, i.e. $N=8$. 
The last variable of the decision vector is a real number which allows the pulse fluence to vary between two predetermined values.
We set
\begin{equation}\label{eq:xn}
X_N = \frac{U}{U_0},
\end{equation}
where $U_0$ is the maximum attainable pulse fluence. 
Bounds on $\vec{X}$ are fixed as
\begin{equation}
\begin{aligned}
\min(X_n) & = 
	\begin{cases} 
	0,  &n < N \\
	\frac{1}{3}, &n = N
	\end{cases}
\\
\max(X_n) & = 1.
\end{aligned}
\end{equation}
The harmonic superposition, \eqref{eq:temporal}, is re-normalized at every objective function evaluation such that the applied THz field has a fluence $U$ which is dictated by the value of $\vec{X}$, i.e. \eqref{eq:xn}.
The lower bound on the fluence is a necessary constraint in the case of minimization over a momentum window since the DE solver must be prevented from converging to fluence values for which the induced carrier density is trivially equal to zero.
We however allow the lower bound to be relaxed in order to provide paths to optimal solutions in the search space, although final pulses are all characterized by $U/U_0 \geq 1/3$.

\begin{figure}
	\includegraphics[width=0.5\textwidth]{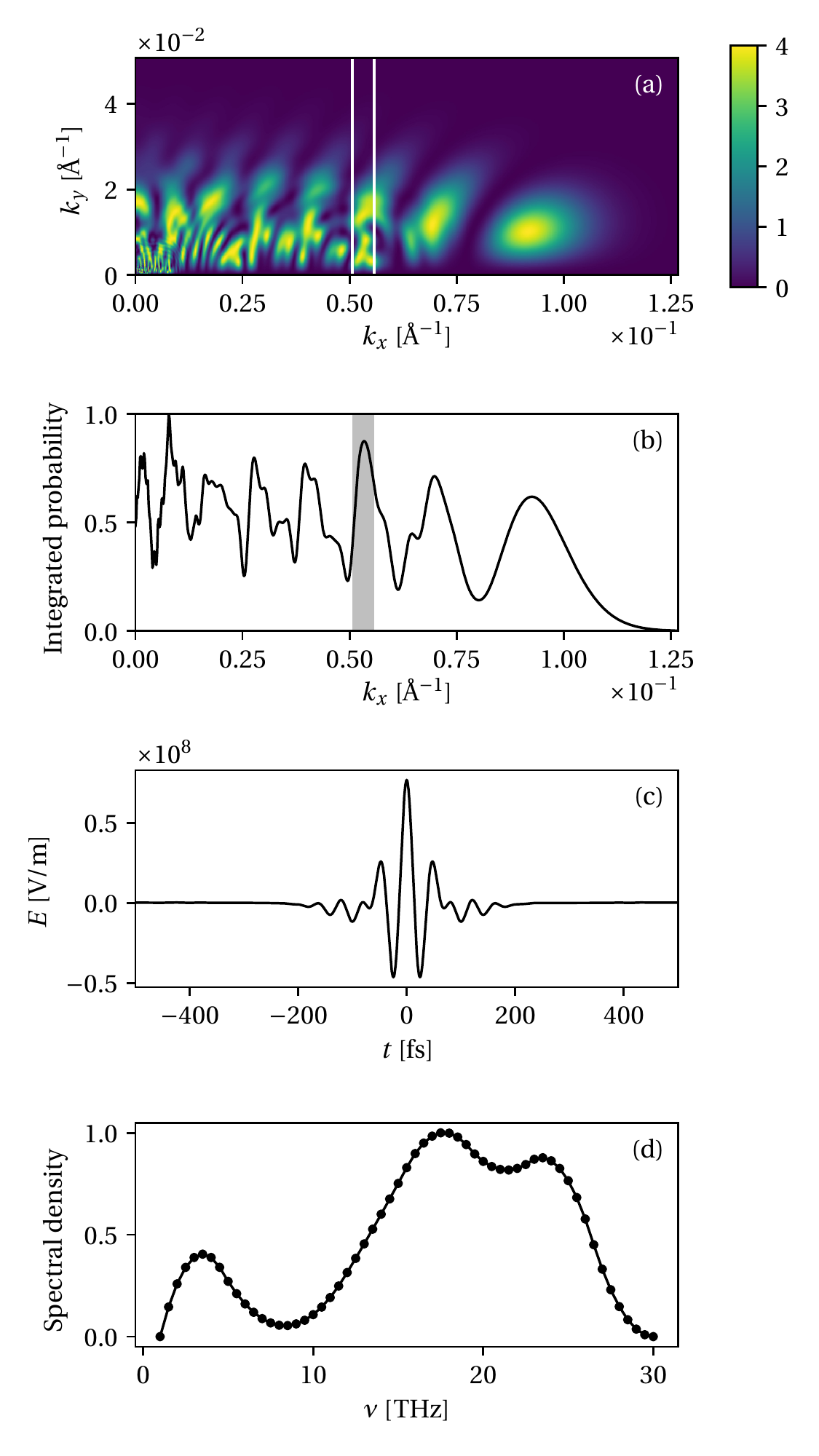}
	\caption{Best found solution for selective maximization of the carrier density.
		(a) Optimized density in momentum space, with the maximization window indicated by a rectangle. (b) Momentum-integrated density, in arbitrary units, showing enhancement in the shaded region. (c) Associated temporal pulse shape. The peak field is determined by the pulse fluence, $U = 2.01 \times 10^{-5}$ J/cm$^2$ . (d) Optimized spectral density from which the temporal shape is obtained (arbitrary units).}
	\label{fig:maximization_fluence}
\end{figure}
\begin{figure}
	\includegraphics[width=0.5\textwidth]{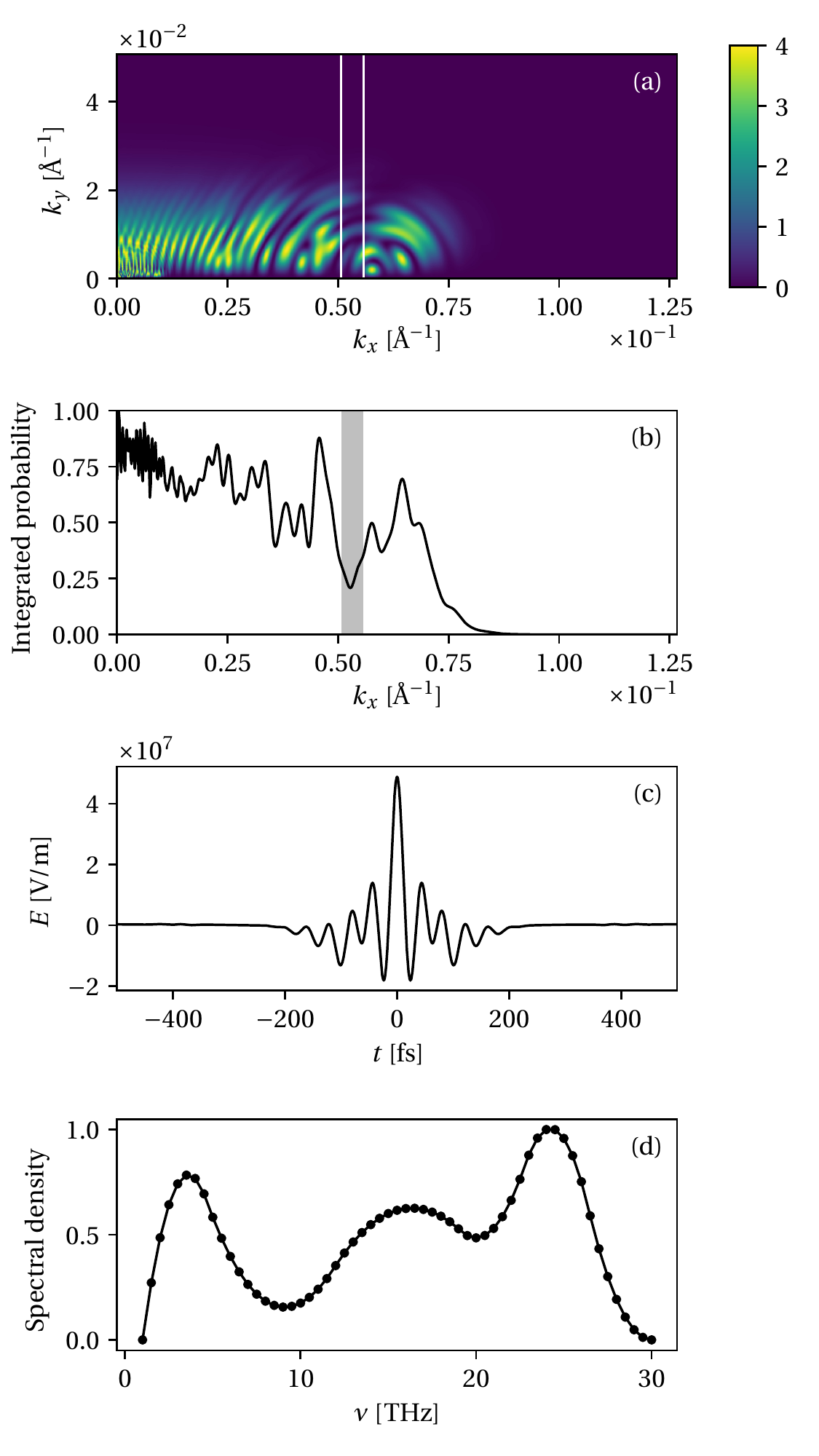}
	\caption{Best found solution for selective minimization of the carrier density.
		(a) Optimized density in momentum space, with the maximization window indicated by a rectangle. (b) Momentum-integrated density, in arbitrary units, showing suppression in the shaded region. (c) Associated temporal pulse shape. The peak field is determined by the pulse fluence, $U = 6.94 \times 10^{-6}$ J/cm$^2$. (d) Optimized spectral density from which the temporal shape is obtained (arbitrary units).}
	\label{fig:minimization_fluence}
\end{figure}

The narrow reciprocal space window over which optimization takes place is chosen as $\mathcal{D}_{\mathbf{p}} = \lbrace 5.07 \times 10^{-2} \mbox{ \AA}^{-1} \leq k_x \leq 5.58 \times 10^{-2} \mbox{ \AA}^{-1},  0 \leq k_y \leq 7.60 \times 10^{-2} \mbox{ \AA}^{-1} \rbrace$.
The maximum pulse fluence is fixed to $U_0 = 2.057 \times 10^{-5}$ J/cm$^2$, meaning that optimized configurations are characterized by a fluence varying between $U_0/3 = 6.86 \times 10^{-6}$ J/cm$^2$ and $U_0$.
Similar to the case of fixed fluence, maximization is carried out via DE for 200 generations using a randomly initialized population of 10 individuals [see Fig. \ref{fig:maximization_fluence}].
This calculation is performed with 4 different random population seeds in order to sample a wider search space. 
A minimization calculation is repeated with the exact same parameters [see Fig. \ref{fig:minimization_fluence}].

Let us examine in more detail the pulse shapes resulting in maximization/minimization of the carrier density over narrow bands in reciprocal space.
The strength of the selective suppression/enhancement is manifest from the momentum integrated spectrum shown in Figs. \ref{fig:maximization_fluence}b and \ref{fig:minimization_fluence}b, which show a peak-to-valley ratio of 2 between suppressed/enhanced peaks and immediately adjacent peaks or valleys.
As can be seen by comparing Figs. \ref{fig:maximization_fluence}c and \ref{fig:minimization_fluence}c, the pulse shapes that result in either minimization or maximization are characterized by a peak field of the same order of magnitude ($\sim 5-7 \times 10^7$ V/m), although the fluence of the pulses which maximizes the carrier density over the selected range is $\sim 3$ times higher than the pulse which minimizes ($2.01 \times 10^{-5}$ J/cm$^2$ and $6.94 \times 10^{-6}$ J/cm$^2$).
This result suggests that it is actually the interplay between spectral components of the pulse, and not merely the peak value of the field, which provides a control knob over the closing or opening of a dynamical gap between the valence and the conduction band of graphene, with a closing leading to CDT, as described in Refs. \cite{Gagnon2017, Gagnon2017b}.

The impact of the spectral content of the pulse can be seen by inspecting Figs. \ref{fig:maximization_fluence}d and \ref{fig:minimization_fluence}d.
In the case of maximization, a relatively broad two-peak structure can be discerned around 20 THz [see Fig. \ref{fig:maximization_fluence}d], whereas in the case of minimization, a narrower peak is apparent around 25 THz in the spectrum [see Fig. \ref{fig:maximization_fluence}d].
This result seems to suggest that in the case of enhancement, a broad spectrum (short pulses) is favored, whereas in the case of suppression, a narrower spectrum (longer pulses) is favored.
This finding is in line with the results obtained in Sec. \ref{sec:cfluence} in the case of maximization over the whole reciprocal space for fixed fluence.
This result should however be taken with care since, as described in Sec. \ref{sec:cfluence}, pulse duration should be as short as possible to prevent carrier cooling from taking place.

\section{Summary}\label{sec:summary}

In this article, we have used differential evolution (DE) to find  THz pulse shapes which result in suppression or enhancement of the laser-induced carrier density in graphene.
Besides providing empirical pulse shapes suited for a specific reciprocal space target, the use of the DE solver enables one to find general trends which should be considered in pulse shaping experiments with Dirac materials.
For example, we found that shorter pulses are generally best suited for maximization of the carrier density, whereas minimization is associated with longer pulse durations.
We also showed that it is possible to vary the photo-induced carrier density by a factor of 4 for a fixed pulse fluence around $1.5 \times 10^{-4}$ J/cm$^2$, using only spectral shape parameters as decision variables of the optimization solver.
This result hints at the fact that THz pulse shaping provides an additional control knob over carrier scattering in graphene \cite{Deffner2015}.
This work may also stimulate the comparison of evolutionary algorithms for pulse shaping problems with established approaches which include numerical ensemble control \cite{Li2006, Li2009, Li2011, Leghtas2011, Chittaro2017}, and emergent tools such as machine learning \cite{Wigley2016, Krefl2017}.

	The authors acknowledge C. Lefebvre for useful discussions and P. Blain for code development.
	D. Gagnon is supported by a  postdoctoral research scholarship from Fonds de recherche du Qu\'ebec -- Nature et technologies (FRQNT).
	Computations were made in part on the supercomputer \emph{Mammouth} from Universit\'e de Sherbrooke, managed by Calcul Qu\'ebec and Compute Canada.
	The operation of this supercomputer is funded by the Canada Foundation for Innovation (CFI), minist\`ere de l'\'{E}conomie, de la Science et de l'Innovation du Qu\'{e}bec (MESI) and FRQNT.

\bibliography{extracted}

\end{document}